\documentclass[prl, reprint, amsmath,amssymb,showpacs,floatfix,longbibliography, aps, twocolumn, superscriptaddress,nofootinbib]{revtex4-1}
\usepackage{braket}
\usepackage{times} 

\usepackage{amsfonts,amssymb,amsmath}
\usepackage[T1]{fontenc}

\usepackage{xr-hyper}
\usepackage{hyperref}
\usepackage{enumitem}
\usepackage{comment}
\usepackage{makecell}
\usepackage[capitalize]{cleveref}
\usepackage{booktabs}

\makeatletter
\newcommand*{\addFileDependency}[1]{
  \typeout{(#1)}
  \@addtofilelist{#1}
  \IfFileExists{#1}{}{\typeout{No file #1.}}
}
\makeatother

\newcommand*{\myexternaldocument}[1]{%
    \externaldocument{#1}%
    \addFileDependency{#1.tex}%
    \addFileDependency{#1.aux}%
}
\myexternaldocument{supplemental}

\usepackage{amsthm}
\newtheorem*{theorem}{Theorem}

\hypersetup{colorlinks=true,citecolor=blue,linkcolor=blue, urlcolor=blue}
\hypersetup{linktocpage}
\newcolumntype{M}[1]{>{\centering\arraybackslash}m{#1}}
\newcolumntype{N}{@{}m{0pt}@{}}
\usepackage{environ}

\bibpunct{[}{]}{,}{n}{}{}

\usepackage{tikz}
\usetikzlibrary{arrows,snakes,backgrounds}

\NewEnviron{eqs}{%
\begin{equation}\begin{split}
    \BODY
\end{split}\end{equation}
}
\newcommand{\Yijia}[1]{ { \color{green} (Yijia: {#1}) }}

\begin{document}
\title{
Boson Sampling for Generalized Bosons
}

\author{En-Jui Kuo}
\thanks{These authors contributed equally to this work.}
\affiliation {Joint Quantum Institute, Department of Physics, NIST and University of Maryland, College Park, Maryland 20742, USA}
\affiliation{Joint Center for Quantum Information and Computer Science, NIST and University of Maryland, College Park,
Maryland 20742, USA}

\author{Yijia Xu}
\thanks{These authors contributed equally to this work.}
\affiliation {Joint Quantum Institute, Department of Physics, NIST and University of Maryland, College Park, Maryland 20742, USA}
\affiliation{Joint Center for Quantum Information and Computer Science, NIST and University of Maryland, College Park,
Maryland 20742, USA}
\affiliation{Institute for Physical Science and Technology, University of Maryland, College Park, Maryland 20742, USA}

\author{Dominik Hangleiter}
\affiliation{Joint Center for Quantum Information and Computer Science, NIST and University of Maryland, College Park,
Maryland 20742, USA}

\author{Andrey Grankin}
\affiliation {Joint Quantum Institute, Department of Physics, NIST and University of Maryland, College Park, Maryland 20742, USA}

\author{Mohammad Hafezi}
\affiliation {Joint Quantum Institute, Department of Physics, NIST and University of Maryland, College Park, Maryland 20742, USA}
\affiliation{Joint Center for Quantum Information and Computer Science, NIST and University of Maryland, College Park,
Maryland 20742, USA}
\date{\today}

\begin{abstract}
    We introduce the notion of ``generalized bosons'' whose exchange statistics resemble those of bosons, but the local bosonic commutator  $[a,a^\dagger]=1$  is replaced by an arbitrary single-mode operator that is diagonal in the generalized Fock basis. 
    Examples of generalized bosons include boson pairs and spins.  
    We consider the analogue of the boson sampling task for these particles and observe that its output probabilities are still given by permanents, so that the results regarding hardness of sampling directly carry over. 
    Finally, we propose implementations of generalized boson sampling in circuit-QED and ion-trap platforms. 
\end{abstract}
\maketitle

Quantum random sampling protocols allow us to demonstrate an advantage of quantum computational devices over classical computers \cite{google_rcs,zhu_quantum_2022,zhong_phase-programmable_2021}. 
In a quantum random sampling protocol the task is to sample from the output distribution of certain random quantum computation.
Surprisingly, even if those computations are not universal, the sampling task can in many cases be computationally difficult for classical computers \cite{aaronson2011computational,hamilton2017gaussian,kruse2019detailed,deshpande_quantum_2022-1,bremner_average-case_2015,boixo_characterizing_2018,bouland_noise_2021,kondo_quantum_2021,gao_quantum_2017,morimae_hardness_2017,bermejo-vega_architectures_2018,haferkamp_closing_2020,bouland_complexity_2018-1,yoganathan_quantum_2019,fefferman2017exact}. 

This is the case even for random linear-optical computations: 
In \emph{boson sampling} \cite{aaronson2011computational} a uniformly random linear mode transformation is applied to a multi-mode bosonic input state and measured in the photon-number basis. 
Boson sampling protocols come in many different variants, ranging from the original proposal of \textcite{aaronson2011computational} with Fock-state input states (FBS), to Gaussian boson sampling (GBS) with Gaussian input states \cite{lund_boson_2014,hamilton2017gaussian,kruse2019detailed} and GBS with threshold detectors \citep{quesada2018gaussian}.
The hardness of simulating those schemes can be traced back to the hardness of computing their output probabilities, which are given by certain polynomials in submatrices of the linear-optical unitary \citep{valiant1979complexity,scheel2004permanents,aaronson2011computational,bjorklund2019faster,kruse2019detailed,deshpande_quantum_2022-1}. 
Importantly, the discovery of GBS has enabled recent experimental demonstrations \cite{zhong_phase-programmable_2021,zhong_quantum_2020} on much larger scales than is possible for FBS \cite{wang_boson_2019} due to the experimental difficulty of Fock state preparation.

In this Letter, we further extend boson sampling protocols to a wider class of quantum systems. 
Specifically, we introduce \emph{generalized bosons}, which is a wide class of particles that hold bosonic commutation relations between different sites but have non-bosonic local commutation relations.
For generalized bosons, the local standard bosonic 
commutation relations are replaced by an arbitrary diagonal operator in the local Fock basis. 
A natural question that we address in this Letter is therefore whether quantum advantage can be demonstrated using those generalized bosonic modes analogously to standard bosons.

Specific instances of generalized bosons were first introduced in Refs.\ \citep{sun1989q, biedenharn1989quantum, rouabah2012q} and find applications for solving integrable systems \citep{kundu2007q, jarvis2016quantum, palazzo2019cylindric, van2016completeness} and even interacting bosonic systems via perturbation theory \citep{tuszynski1993statistical}. 
As we discuss in more detail below,
modes obeying generalized commutation relation can also be found in AMO systems. 
While standard bosons are non-interacting, systems with non-trivial diagonal commutation relations can be viewed as interacting. 
Examples include conventional spin degrees of freedom, and the so-called paraboson \citep{jagannathan1981boson,chaturvedi1991bose} that has recently been studied in ion-trap systems \cite{linke_paraboson}. 
Below, we present and analyze another variant of generalized bosons in a circuit-QED setup taking the form of boson pairs \cite{PhysRevLett.123.063602,symmetry_breaking_pair,Yale2015_twophoton_loss,puri2017_twophoton,touzard2018_twophoton}.

\begin{figure}[t]
\begin{centering}
\includegraphics[scale=0.40]{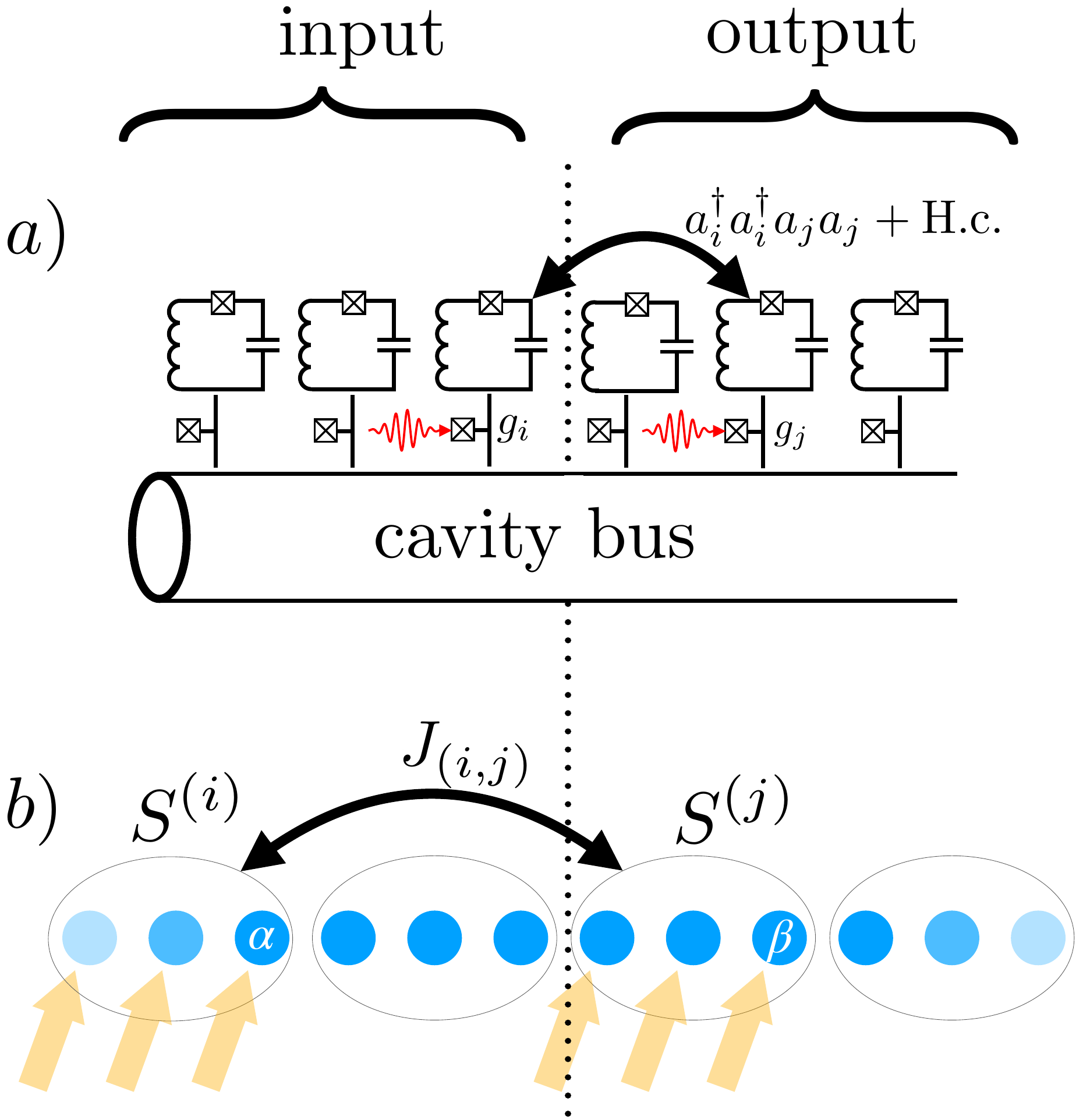} 
\par\end{centering}
\caption{
Boson sampling with generalized bosons. (a) Circuit-QED implementation for boson pairs $b_{i}=a_{i}^{2}$
as cavity excitations. The mode-mixing operation is implemented as a non-linear photon-pair tunneling. (b) Trapped-ion implementation of super-spins. Generalized boson (superspin) is encoded in internal atomic states.  As described in the text the mode-mixing unitary operation
is implemented in a Trotterized way by a sequential application of
Molmer-Sorensen \cite{molmer1999multiparticle} laser beams to
each pair of ions. In both cases, the final step is to perform local population measurement.}
\label{Fig}
\end{figure}

Within this framework,
we show that boson sampling can be simulated efficiently  by generalized bosons using Fock state preparations, occupation number measurements and linear mode mixing. 
Consequently, all the complexity results for the  original boson sampling protocol carry over. 
While linear mode mixing is naturally implemented in non-interacting systems only, we re-interpret a result by \textcite{peropadre2017equivalence} to show an approximate, but efficient, simulation of mode-mixing for generalized bosons, in certain limits. 
Finally, we provide specific implementation proposals for a circuit-QED and an ion-trap platforms. 

On a high level, our work can be viewed as addressing the question whether there is an intermediate system in between qubits and non-interacting standard bosons in terms of the level of interaction.
Our results show that performing mode mixing transformations ``bypasses'' the interactions of generalized bosonic systems, giving rise to the same output probabilities. 

\paragraph{Generalized bosons---}
Let us start by being more specific about the definition of generalized bosonic particles. 
Recall the standard bosonic commutation relations between bosonic annihilation and creation operators $a_i$ and $ a_i^\dagger$ for a mode $i$
\begin{equation}
[a_{i}^{\,},a_{j}^{\dagger}]=\delta_{ij},\quad
[a_{i}^{\dagger},a_{j}^{\dagger}]=[a_{i}^{\,},a_{j}^{\,}]=0,\label{eq:comm_rel}
\end{equation}
where ${[\ \cdot\ ,\ \cdot\ ]}$ denotes the commutator and $\delta_{ij}$
is the Kronecker delta.

For generalized bosons, the last two commutation relations in \cref{eq:comm_rel} remain unchanged, while the first commutation relation
is modified by multiplying the Kronecker delta by an arbitrary diagonal operator characterized by a  function $F: \mathbb N \rightarrow \mathbb C$ of the single-mode occupation number. 
Specifically, we define generalized bosonic operators $b_i, b_i^\dagger$ annihilating and creating an excitation in mode $i$, respectively, by their commutation relations
\begin{align}
    &  [b_{i}^{\,},b_{j}^{\dagger}]=\delta_{i,j} \sum_{n_i=0}^{\infty}F(n_i)\ket{n_i}\bra{n_i}, \quad
    [b_{i}^{\dagger},b_{j}^{\dagger}]=[b_{i}^{\,},b_{j}^{\,}]=0\label{eq:comm_rel}.
\end{align}
Here, the generalized Fock state $\ket{n_i}$ with occuptation number $n_i$ in mode $i$ is defined by the action of the creation operator $b_i^\dagger$ on the vaccum as $(b_i^\dagger)^{n_i} \ket 0 = f(n_i) \ket {n_i}$, where $f : \mathbb N \rightarrow C $ is a function that alternatively characterizes the generalized boson. 
$f$ and $F$ are related and their exact correspondence is discussed in the Supplementary Material \cite[Sec.~I]{suppmaterial}. 
For convenience, we call $f(n)$ the \textit{bosonic factor} and only use $f(n)$ in the following.
A multimode Fock state of generalized bosons on $M$ modes can thus be written as 
\begin{equation}
\ket{n_{1},n_{2},\ldots ,n_{M}}=\left(\prod_{i=1}^{M}\frac{1}{f(n_{i})}\right)b_{1}^{\dagger n_{1}}b_{2}^{\dagger n_{2}} \cdots b_{M}^{\dagger n_{M}}\ket{0}.
\end{equation}
We give some examples of generalized bosons and their corresponding bosonic factors in \cref{t1}.  

The basic idea of the FBS protocol due to \textcite{aaronson2011computational} is to send a Fock state of $N$ photons in $M$ modes into a uniformly random linear mode-mixing circuit described by a unitary matrix $\Lambda \in U(M)$, and subsequently, to measure the output state in the Fock basis. 
The linear optical network $\Lambda$ performs mode-mixing so that the input mode operators $\{a_i\}_{i=1}^M$ are transformed to output operators $\tilde{a}_i = \sum_{j=1}^M \Lambda_{ij} a_j$. 
The probability of obtaining an outcome $\mathbf k = (k_1, \ldots, k_M)$ given that the input configuration is given by $\mathbf l = (l_1, \ldots, l_M)$ for $k_i, l_i \in \mathbb N$ such that $\sum_i k_i = \sum_i l_i = N$ is then given by 
\begin{align}
\mathrm{Pr} (\mathbf k| \mathbf l) = 
\frac{
    \left|\mathrm{Perm} (\Lambda[\mathbf k| \mathbf l]) \right|^2 
} {
    \left( \prod_i l_i!\right) \left(\prod_i k_i!\right) 
}   .
\end{align}
Here, the permanent $\mathrm{Perm}$ of an $N \times N$ matrix $A = (A_{i,j})_{i,j}$ is defined like the determinant but with only positive signs as 
\begin{align}
    \mathrm{Perm}(A) = \sum_{\sigma \in \mathbb S_N} \left(\prod_{i=1}^N A_{i,\sigma(i)}\right), 
\end{align}
where $\mathbb S_N$ is the symmetric group of $N$ elements. 
Moreover, 
$\Lambda[\mathbf m| \mathbf n]$ is an $(N \times N)$ matrix constructed by repeating the $i^{\text{th}}$ column of $\Lambda$ $l_i$ many times, and the $j^{\text{th}}$ row $k_i$ many times.

\begin{table*}[t]
\begin{tabular}{l >{\qquad } l<{\qquad}  l  >{\qquad}l }
    \toprule
    Boson type& Definition  & $F(n)$  & $f(n)$   \\[.5ex] \midrule
Standard boson & $a$ & $1$ & $\sqrt{n!}$ \\[2ex]
Boson pair \cite{PhysRevLett.123.063602,symmetry_breaking_pair,Yale2015_twophoton_loss,puri2017_twophoton,touzard2018_twophoton} & $b=a^2$ &  
$2+8n $
& $\sqrt{(2n)!}$ 
\\[1ex]
Spin-$S$ boson \citep{chryssomalakos2018geometry, radcliffe1971some}  & $b=S_x-iS_y$ & 
$(n-2S) \, \theta(2S-n)  $ & $\left(\frac{n! (2S)!}{(2S-n)!}\right)^{\frac{1}{2}}$ \\[2ex]
$q$-boson \citep{sun1989q, biedenharn1989quantum, rouabah2012q} & $b b^{\dagger}-qb^{\dagger}b=q^{N}, q \in \mathbb{C}$  & $\frac{[n+1]_{q}!}{[n]_{q}!}-\frac{[n]_{q}!}{[n-1]_{q}!}$ & $\sqrt{[n]_{q}!}$ \\[1ex]
$m$-paraboson \citep{jagannathan1981boson,chaturvedi1991bose} & $[b, b^{\dagger}]=1+(2m+1)e^{i \pi N_m}$& $1+(2m+1)e^{i \pi n}$ & $\prod_{k=1}^{n} \left(
    \frac{
        2k+2m+3+(2m+1)e^{2\pi k}
        }{
        2}
        \right)^{\frac 12}$\\
\bottomrule
\end{tabular}
\caption{\label{t1}Various generalized bosons and their generalized bosonic algebra, defined in terms of either the function $F(n)$ or $f(n)$, see \cref{eq:comm_rel}. For spin-$S$ boson, the spin operators are $S_x=\sum_{j=1}^{2S} \sigma_x$, $S_y=\sum_{j=1}^{2S} \sigma_y$, $\theta(x)$ is the Heaviside step function which is zero for $x<0$ and $1$ for $x \geq 0$.
For q-bosons, $N$ is defined such that $[b_{q}, N]=b_{q}, [b_{q}^{\dagger} , N]=-b_{q}^{\dagger}$.  $[n]_q=\frac{q^n-q^{-n}}{q-q^{-1}}$, $[n]_q!=\prod_{j=1}^n [j]_q$,
$N_m$ is the number operator of $m$-paraboson. 
}
\end{table*}

\paragraph{Boson sampling for generalized bosons---}
We are now ready to present the main theoretical result of our work, namely, that the output probabilities of sampling from generalized bosons with Fock-state inputs are proportional to the output probabilities of standard FBS.

\begin{theorem}
\label{thm:fbs}
    Consider a linear transformation $\Lambda \in U(M)$ of $M$ modes of a generalized bosonic algebra on those modes with bosonic factor $f$. 
    Then the probability of measuring outcome $\mathbf k$ given a Fock input state $\ket{\mathbf l}$ is given by 
    \begin{equation}\label{ge-perm}
    \mathrm{Pr}(\mathbf k | \mathbf l) =    \left(\prod_{i=1}^{N} \frac{f(k_i)}{f(l_i) k_i!}\right) ^2 \left| \mathrm{Perm} (\Lambda[\mathbf k| \mathbf l]) \right|^2 .
    \end{equation}
\end{theorem}

We prove the Theorem in the Supplementary Information~\cite[Sec.~II]{suppmaterial}. 
Consequently, the complexity of FBS for generalized boson remains the same as the complexity in the standard boson case \cite{aaronson2011computational}.
As it turns out, it is also possible to construct an analogue of the Gaussian phase-space formalism and a corresponding GBS, which however, is highly unnatural for generalized bosons and hence we defer a detailed discussion to the Supplementary Material~\cite[Sec.~III]{suppmaterial}.

We have pointed out that the output probabilities of FBS are essentially unchanged under varying the diagonal commutation relations between the bosons. 
But does this connection extend beyond being a mathematical curiosity? 
While the preparation of Fock states, the implementation of a linear-optical mode transformation and a measurement in the Fock basis are natural operations in the context of quantum optics, 
whether the same is true for generalized bosons is unclear a priori. Vice versa, our result implies that FBS can be implemented whenever those operations are possible.
For those platforms, it can then be used as a quantum advantage benchmark to compare their performance in a hard-to-simulate regime with other platforms.

\paragraph{General implementation scheme---}
We now describe the main idea of implementing FBS for generalized bosons by generalizing the implementation for a spin-$\frac{1}{2}$ chain due to \textcite{peropadre2017equivalence}.
While Fock-state preparations and measurements are natural, and for every fixed particle number there exists a unitary that realizes linear mode mixing,  
it does not seem possible to implement this unitary efficiently in general. 
Recall that a mode-mixing transformation $\exp(- i t H)$ arises naturally for standard bosons evolved for time $t$ under the quadratic Hamiltonian $H= \sum_{i,j=1}^M h_{i,j} a_i^\dagger a_j$ with coefficient matrix $h \in \mathbb C^{M \times M}$. 
In contrast, the non-trivial commutation relation of generalized bosons create arbitrary higher-order terms in the Baker-Campbell-Hausdorff expansion when evolving an operator $b_i$ under a Hamiltonian that is quadratic in the generalized bosonic operators $b_i, b_i^\dagger$. 
In other words, quadratic Hamiltonians of generalized bosons are interacting, and vice versa, non-interacting evolution is generated by highly complex Hamiltonians. 

The idea of \textcite{peropadre2017equivalence} is to prevent those interactions between individual bosons from happening. 
To achieve this, they perform a space-time mapping, allowing them to swap input modes with output modes in a single oscillation so that there are no collisions during the time evolution. 
Specifically, they consider a system of $2M$ modes separated into $M$ input and $M$ output modes, which is evolved for a short, constant time under  the Hamiltonian
\begin{align}
\label{eq:peropadre Hamiltonian}
H_{\rm BS} = \sum_{i,j=1}^M  (a_{j,out}^{\dagger} R_{ji} a_{i,in} + \mathrm{h.c.})
\end{align}
acting on standard bosonic modes with annihilation operators $\{a_{i,in}\}_{i=1}^M$ and $\{a_{i,out}\}_{i=1}^M$ on the two halves of the system, respectively.
Here $R \in U(M)$ is a unitary matrix. 
It turns out that evolving the initial Fock state $\ket{\phi(0)} = \ket{\mathbf l^M}_{in} \otimes \ket{0^M}_{out}$ 
under this Hamiltonian for time $\pi/2$ results precisely in a state 
\begin{align}
\label{eq:peropadre time evolved state}
\ket{\phi(\pi/2)} = (-i)^N \prod_{i=1}^N \sum_{j=1}^M \overline R_{j,i} a_{j,out}^\dagger  \ket 0 , 
\end{align}
so that a measurement in the Fock basis on the output modes reproduces the boson sampling protocol. 

Let us now consider the analogous generalized bosonic Hamiltonian
\begin{align}
\label{eq:gen bos hamiltonian}
H_{\rm gBS} = \sum_{i,j=1}^{M} b_{j,out}^{\dagger}R_{ji} b_{i,in} +\text{h.c}  
\end{align}
evolved for time $\pi/2$ with initial state $\ket{\phi(0)}$ conceived of as a generalized Fock state. 
Then the generalized bosonic algebra will introduce an error into the output state, since during the time evolution bosons will `meet' and thus experience the non-trivial diagonal commutation relation. 

\textcite{peropadre2017equivalence} show that for spin-$1/2$ systems, in the dilute limit of $M \in \Omega(N^4)$ and for initial states with only $0$ or $1$ particles in a mode will go to zero as $O(N^2/\sqrt{M})$ in Frobenius norm in the asymptotic limit. 
Importantly, this translates into a total-variation distance bound between the corresponding output distributions of $O(N^2/\sqrt M)$.
Hence, their result can be read as showing that in this regime, the Hamiltonian \eqref{eq:gen bos hamiltonian} approximately realizes linear mode mixing under the unitary~$R$. 

In the following, we devise two possible implementations of the generalized boson sampling protocol by simulating the Hamiltonian \eqref{eq:gen bos hamiltonian} in circuit-QED and trapped-ion platforms, respectively. 
Each protocol includes the preparation of a Fock-state of generalized bosons, a simulation of Hamiltonian time evolution under $H_{\rm gBS}$, and a measurement in the Fock basis of generalized bosons.

\paragraph{Circuit-QED implementation---}

We start by considering photon pairs as the generalized boson where
the annihilation operator is simply the square of a standard boson
annihilation operator: $b_{i}\equiv a_{i}^{2}$ with $[a_{i},a_{j}^{\dagger}]=\delta_{i,j}$.
It is straightforward to show that $[b_{i},b_{j}^{\dagger}]=\sum_{n_i=0}^{\infty}(2+8n_i)\ket{n_i}\bra{n_i}\delta_{ij}$. This expression is clearly
diagonal in the Fock basis as required by the definition of generalized
bosons; see also \cref{t1}. 
Although photon pairs have been generated in various parts
of the electromagnetic spectrum, our scheme requires mode-mixing of
photon pairs that is \textit{quartic} in the standard bosonic operators. 
Fortunately,
the generation and manipulation of such photon pairs have been extensively
studied in circuit-QED systems \citep{PhysRevLett.123.063602,symmetry_breaking_pair,Yale2015_twophoton_loss,puri2017_twophoton,touzard2018_twophoton,ma2021_twophoton}.
In particular, we consider an array of non-linear resonator interacting
with a bus waveguide via a non-linear process (see \cref{Fig}(a)) as 
described by the Hamiltonian
\begin{align}
H_{{\rm eff}} & =\sum_{i}g_{i}a_{i}^{\dagger2}c+\text{H.c.}-\Delta c^{\dagger}c-\chi\sum_{i}a_{i}^{\dagger}a_{i}^{\dagger}a_{i}a_{i}\label{eq:H_cqed}
\end{align}
 in the frame rotating at the cavity
frequency $\omega_{c}$. 
Here, $\chi,g_{i}$ and $\Delta$ are the circuit parameters defined
in the supplementary material, $a_{i}$ is the cavity annihilation
operator and $c$ is the bus bosonic resonator mode annihilation operator satisfying $[c,c^\dagger]=1$. In particular, we assume that the coupling $g$ can be externally controlled.
We note that our implementation allows for the dynamical independent laser control of
coupling coefficients $g_{i}=g_{i}(t)$. By adiabatically eliminating
the bus cavity mode, we find the effective Hamiltonian $H_{\text{pair}}=\sum J_{i,j}\left(t\right)b_{i}^{\dagger}b_{j}$
with the tunneling being $J_{i,j}=g_{i}(t)g_{j}^{*}(t)/\Delta-\chi\delta_{i,j}$.
We note that the on-site
interaction $\chi$ will be ignored in the following as it can be
eliminated by either choosing the coupling constants such that $|g_{i}|^{2}/\Delta=\chi$
or by coupling to an additional Josephson junction in analogy to \citep{PhysRevLett.123.063602}. 
The Hamiltonian $H_{\text{pair}}$ is formally equivalent to $H_{\rm gBS}$
required to perform the mode-mixing operation \cite{peropadre2017equivalence}. 
In the case in which the coefficients $g_i$ are time-independent, the resulting $J_{i,j}$ is only a rank-1 matrix and therefore does not allow for the implementation of an arbitrary unitary mode transformation. 
A more general interaction pattern can be obtained by, e.g., using a multimode cavity. 
Alternatively, due to the independent dynamical control over coupling terms $g_i$ any Hamiltonian can be implemented in a Trotterized fashion as shown in the Supplementary Material \cite[Sec.~V]{suppmaterial}. 

We now discuss the  Fock-state $b^\dagger \ket 0$ preparation step of the protocol.
Arbitrary Fock-state preparation has been demonstrated experimentally in  circuit QED setups \cite{Hofheinz:2008wb}.
For that we first assume the non-interacting limit $g_{i}\rightarrow 0$. In this case
in the non-rotating frame the Hamiltonian \cref{eq:H_cqed}
is local and diagonal in the Fock basis $\ket{n_{i}}$ with eigenvalues  $\varepsilon_{n}=n\omega_{c}-n(n-1)\chi$. In order to prepare
the desired Fock state we now assume the system is initially prepared
in the vacuum state $\ket 0$. The system is then weakly
driven with the driving Hamiltonian $H_{\text{dr}}=\Omega_{\text{dr}}/2\sum_{i}(a_{i}\exp(i\omega_{\text{dr}}t)+a_{i}^{\dagger}\exp(-i\omega_{\text{dr}}t))$,
where $\omega_{\text{dr}}$ and $\Omega_{\text{dr}}$ are the frequency
and the amplitude of the driving. In the limit of weak driving $\Omega_{\text{dr}}\ll\chi$,
upon tuning the driving frequency to the two-photon resonance $\omega_{\text{dr}}=\varepsilon_{2}/2$,
the system undergoes a two-photon Rabi oscillation $\ket {0_i} \rightarrow b_i^\dagger \ket 0 $
with the period given by a two-photon Rabi frequency $\Omega_{2}=\sqrt{2}\Omega_{\text{dr}}^{2}/(4\chi)$.

The leading error in Fock state preparation is due to populating the state $|1\rangle$ with probability
$p_{1}\leq \Omega_{\text{dr}}^{2}/(4|\varepsilon_{1}-\omega_{\text{dr}}|^{2})$.  This probability can be minimized by reducing the Rabi frequency, trading off against decoherence. 
A further improvement may be achieved by employing the adiabatic protocol~\cite{PhysRevA.100.033822}, which assumes control over the detuning of the drive. 

The final step of the protocol is the measurement of the photon number
distribution. 
This can be done by standard means, e.g., by employing the
quantum nondemolition measurement protocol as experimentally demonstrated
in \cite{schuster2007resolving,PhysRevX.10.021060}. The fidelity of the measurement is limited by the cavity decoherence. 

\paragraph{Trapped-ion implementation---} 
We now discuss implementation of spin-$S$ generalized bosons  
using a trapped-ion quantum simulator, as schematically shown in Fig.~\ref{Fig}(b). To this end, we consider a chain of ions in a linear Paul trap \cite{raizen1992ionic}. 
Each ion is considered to be a two-level system $\{\left|g\right\rangle _{\alpha},\left|e\right\rangle _{\alpha}\}$
with the corresponding transition frequency denoted as $\omega_{eg}$.
We encode the super-spin $S^{(i)}$ as collective excitation of a
subset of ions $\left\{ \alpha\right\} _{i}$ such that the lowering
operator can be defined as: $S_{\pm}^{\left(i\right)}\equiv N^{-1/2}\sum_{\{\alpha\}_{i}}\sigma_{\pm}^{(\alpha)},$
where $N$ is the number of ions encoding the superspin and $\sigma_{-}^{(\alpha)}\equiv\left|g\right\rangle _{\alpha}\left\langle e\right|$,
$\sigma_{+}^{(\alpha)}\equiv\left|e\right\rangle _{\alpha}\left\langle g\right|$.

We split our implementation into three steps: state preparation, unitary mode mixing via \cref{eq:peropadre time evolved state}, and measurement. 
The initial superspin Fock (Dicke)-state preparation can be performed using the technique experimentally demonstrated in Ref.~\cite{hume2009preparation}. It consists of two steps: first the preparation of the motional Fock state by selectively driving the first motional blue sideband. The second step consists of driving the target superspin ions resonantly with the red motional sideband thereby transferring the excitation into the Dicke state $N^{-1/2}\sum_{\{\alpha\}_i}\vert e\rangle_{\alpha}$.

We now discuss a Trotterized way \cite{lanyon2011universal} of implementing time evolution under the long-range Heisenberg-exchange type interaction $H_{\text{int}}=J_{(i,j)}S_{+}^{(i)}S_{-}^{(j)}+\text{H.c.}$ featuring in the Hamiltonian \eqref{eq:peropadre Hamiltonian}.
As extensively discussed in the literature \cite{brydges2019probing},
such an interaction between any pair of ions
can be generated driving ions in a generalized M{\o}lmer-Sorensen \cite{molmer1999multiparticle}, with tailored 
laser configuration \cite{davoudi2020towards}. More precisely, we consider a pair of bichromatic
laser beams driving the transition $e\leftrightarrow g$ with the
laser frequencies respectively tuned to $\omega_{eg}+\Delta$ and
$\omega_{eg}-\Delta$ where $\Delta$ is laser detuning. As we discuss in the Supplementary Material \cite[Sec.~IV]{suppmaterial} this
generates the ion-ion interaction characterized by the Hamiltonian
$H_{(\alpha,\beta)}^{(XY)}\approx J^{(\alpha,\beta)}\{\sigma_{+}^{(\alpha)}\sigma_{-}^{(\beta)}+\text{H.c.}\}$, 
where the interaction coefficient scales as $J_{(\alpha,\beta)}=J_{0}/\left|\alpha-\beta\right|^{\zeta}$
with $0<\zeta<3$ and $J_{0}$ is interaction constant. Time evolution under the desired
Hamiltonian $H_{\text{int}}$ can then be obtained in a Trotterized way
by running the interaction $H_{\alpha,\beta}^{(XY)}$ for each pair of ions
for a time $\delta t_{\alpha,\beta}$ inverse proportional to the interaction
strength $J_{(\alpha,\beta)}$ to preserve the spin symmetry $\delta t_{i,j}^{\alpha,\beta}\approx\delta t_{i,j}\times\tilde{J}/J_{\alpha,\beta}$,
where $\tilde{J}=\text{min}_{\{i\}_{\alpha}\{j\}_{\beta}}\left[J_{i,j}\right]$.
To the lowest order in Floquet-Magnus expansion we find:

\begin{align}H_{\text{eff}} & \approx J_{0}N\frac{\sum_{i,j}\delta t_{i,j}\left\{ S_{+}^{(i)}S_{-}^{(j)}+\text{H.c.}\right\} }{\sum_{i,j}\sum_{\alpha,\beta}\delta t_{i,j}\times\left|\alpha-\beta\right|^{\zeta}}.\label{eq:Heff}\end{align}

By performing the summation and assuming the densest unitary operation
($\delta t_{i,j}=\delta t$) for $\zeta\approx0$ and $N=2$, corresponding to $S=1$,
we find the interaction between $N_{S}$ superspins to be of the order $\text{max}(J^{(\alpha,\beta)})=4\times J_{0}/\left(N_{S}^{2}-N_{S}\right)$.
We note that this scaling can potentially be improved by performing optimization of the M{\o}lmer-Sorensen laser configuration \cite{teoh2020machine}.
The number of available superspins can be estimated as $\text{max}(J^{(\alpha,\beta)})>\gamma$,
where the decoherence rate is $\gamma \approx1\text{Hz}$. 
By taking $J_{0}\approx 1 \text{kHz}$
we find the number of available superspins to be of the order of $N_{S}\approx 50$.

Measurement results in
the collective basis can be inferred from just local spin measurements
due to the restriction to the collective states of each superspin.

\paragraph{Outlook---}
Universal circuit sampling and boson sampling formalize natural notions of random computations in the circuit model and the linear-optical model of computation, respectively.
Viewed from the perspective of generalized bosons, these systems are captured by spins and standard bosons, respectively. 
In contrast to standard bosons 
spins are strongly interacting. 
Our results can be viewed as addressing the question of whether there is an intermediate system between qubits and non-interacting standard bosons in terms of the level of interaction.
Indeed, generalized bosons provide a natural framework for thinking about this question and we make some first progress by showing that boson sampling can be simulated in such intermediate systems.
Having said that, it is less clear that this is indeed the most natural notion of random computation in these contexts. 
An exciting open question is thus to identify as well as assess the computational complexity of natural random computations for various physical platforms.

A key open question for generalized boson sampling is the question of whether it is possible to certify samples produced in such models. 
For standard bosons, state preparations can indeed by verified by making use of the formalism of Gaussian quantum information \cite{chabaud_efficient_2021}. 
As we discuss in more detail in the Supplementary Material \cite{suppmaterial}, this formalism does not carry over to generalized bosons, and it is therefore an interesting open to show how generalized bosonic state preparations can be efficiently verified.
 \\

\paragraph{Acknowledgements.} We thank Ze-Pei Cian, Abhinav Deshpande, and Yixu Wang for useful discussions. 
This work was supported by ARO W911NF-15-1-0397, National Science Foundation QLCI grant OMA-2120757, AFOSR-MURI FA9550-19-1-0399, Department of Energy QSA program, and the Simons Foundation.
D.H.\ acknowledges financial support from the U.S.\ Department of Defense through a QuICS Hartree fellowship. M.H. thanks ETH Zurich for the hospitality during the conclusion of this work. 

\bibliography{references.bib}

\end{document}


\title{SUPPLEMENTAL MATERIAL}

\author{En-Jui Kuo}
\thanks{These authors contributed equally to this work}
\affiliation {Joint Quantum Institute, Department of Physics, NIST and University of Maryland, College Park, Maryland 20742, USA}
\affiliation{Joint Center for Quantum Information and Computer Science, University of Maryland, College Park,
Maryland 20742, USA}

\author{Yijia Xu}
\thanks{These authors contributed equally to this work}
\affiliation {Joint Quantum Institute, Department of Physics, NIST and University of Maryland, College Park, Maryland 20742, USA}
\affiliation{Joint Center for Quantum Information and Computer Science, University of Maryland, College Park,
Maryland 20742, USA}
\affiliation{Institute for Physical Science and Technology, University of Maryland, College Park, Maryland 20742, USA}

\author{Dominik Hangleiter}
\affiliation{Joint Center for Quantum Information and Computer Science, University of Maryland, College Park,
Maryland 20742, USA}

\author{Andrey Grankin}
\affiliation {Joint Quantum Institute, Department of Physics, NIST and University of Maryland, College Park, Maryland 20742, USA}

\author{Mohammad Hafezi}
\affiliation {Joint Quantum Institute, Department of Physics, NIST and University of Maryland, College Park, Maryland 20742, USA}
\affiliation{Joint Center for Quantum Information and Computer Science, University of Maryland, College Park,
Maryland 20742, USA}

\maketitle

\tableofcontents

\section{Generalized Bosons\label{SM:generalized_FBS}}
In this section, we introduce the definition of generalized boson. In the next section, we will then show that FBS will still yield a permanent when sampling from the output distribution of linear mode mixing applied to  generalized Fock states.

To define generalized bosons, first, recall the definition of a standard boson from regular quantum field theory \citep{peskin2018introduction}. 
Let $A$ be a nonempty set. 
We can associate each $i \in A $ with a pair of operators $a^{\,}_i, a^\dagger_i$, corresponding to annihilating and creating a boson, respectively. 
Those operators satisfy the following commutation relations
\begin{equation}
[a^{\,}_i, a^\dagger_j] \equiv a^{\,}_i a^\dagger_j - a^\dagger_ja^{\,}_i = \delta_{i j},\quad 
[a^\dagger_i, a^\dagger_j] = [a^{\,}_i, a^{\,}_j] = 0,
\end{equation}
where $\delta_{i j}$ is the Kronecker delta. 

For generalized bosons, we proceed analogously, except that we relax the diagonal commutation relation requirement in the following sense 
\begin{equation}
[b_{i}^{\,},b_{j}^{\dagger}]=\delta_{i,j} \sum_{n_i=0}^\infty F(n_i) \ket{n_i} \bra{n_i}, \quad
    [b_{i}^{\dagger},b_{j}^{\dagger}]=[b_{i}^{\,},b_{j}^{\,}]=0,
\end{equation}
where $F: \mathbb N_0 \rightarrow \mathbb C$ may be an arbitrary complex-valued scalar function. We recover the definition of standard bosons by defining $F(n) \equiv 1, \,  \forall n \in \mathbb N_0$, which corresponds to the identity operator.  

We define the generalized Fock basis by the action of the generalized bosonic creation operator on the vacuum state as
\begin{equation}\label{single}
(b^{\dagger})^n \ket{0}=f(n)\ket{n},
\end{equation}
where $f: \mathbb N_0 \rightarrow \mathbb C$ is an alternative characterization of a generalized boson that is equivalent to $F$. For $n=0,$ we always have $f(0)=1$ due to Eq.~\eqref{single}.

To see this, let us first show how $F$ is determined by $f$. 
To this end, observe the following standard relation for the action of a creation and annihilation operator on a Fock state $\ket n $
\begin{equation}
    b^{\dagger}\ket{n}=\frac{f(n+1)}{f(n)}\ket{n+1}, \quad 
    b\ket{n}=\frac{f(n)}{f(n-1)}\ket{n-1}. 
\end{equation}
Given this, we can compute  
\begin{equation}
   F(n) =  \bra{n}[b,b^{\dagger}]\ket{n}=\frac{f(n+1)^2}{f(n)^2}-\frac{f(n)^2}{f(n-1)^2}. 
\end{equation}
Conversely, we can construct $f$ given $F$ recursively via the following relation 
\begin{equation}
\label{eq:recursive f F}
   \frac{f^2(n+1)}{f^2(n)}=\frac{f(1)^2}{f(0)^2}+\sum_{i=1}^{n}F(i). \forall n \geq 1, 
\end{equation}
and the requirement of $f(0) = 1$ as well as the observation that $\frac{f(1)^2}{f(0)^2}=F(0)$ so that $f(1)=\sqrt{F(0)}$.
So once we fixed $F(i)$ $\forall i \in \mathbb{N}$ and $f(0) = 1$, $f$ is recursively determined by Eq.~\eqref{eq:recursive f F}. 
In particular,  if we choose $F(i)=1, \forall i$ we obtain$f(n)=\sqrt{n!}$ as we would expect.

So the bosonic factor completely determines all the structure of generalized bosons and vice versa. 
We will see how the order structure of $f(n)$ will play an important rule in our context. 

We can then define the multimode Fock state
of generalized bosons in $M$ modes as 
\begin{equation}
\ket{n_{1},n_{2},...,n_{M}}=\left(\prod_{i=1}^{N}\frac{1}{f(n_{i})}\right)b_{1}^{\dagger n_{1}}b_{2}^{\dagger n_{2}}...b_{M}^{\dagger n_{N}}\ket{0}^{M}.
\end{equation}

\section{Bosons sampling with generalized bosons}
\label{SM:FBS}

We now derive FBS for generalized boson and show its output probability is still proportional to a permanent of mode-mixing matrix $\Lambda$. 
\begin{theorem}
    Consider a linear transformation $\Lambda \in U(M)$ of $M$ modes of a generalized bosonic algebra on those modes with bosonic factor $f$. 
    Then the probability of measuring outcome $\mathbf k$ given a Fock input state $\ket{\mathbf l}$ is given by 
    \begin{equation}\label{ge-perm}
    \mathrm{Pr}(\mathbf k | \mathbf l) =    \left(\prod_{i=1}^{N} \frac{f(k_i)}{f(l_i) k_i!}\right) ^2 \left| \mathrm{Perm} (\Lambda[\mathbf k| \mathbf l]) \right|^2 .
    \end{equation}
\end{theorem}
where $\Lambda[\mathbf m| \mathbf n]$ is an $(N \times N)$ matrix constructed by repeating the $i^{\text{th}}$ column of $\Lambda$ $l_i$ many times, and the $j^{\text{th}}$ row $k_i$ many times.
\begin{proof}
The unitary  transformation $\Lambda$ acts like the following:
\begin{equation}
\boldsymbol{b} \to \Lambda^{\dagger} \boldsymbol{b}, \boldsymbol{b}^{\dagger}  \to \Lambda^{T} \boldsymbol{b}.
\end{equation}
We can expand the $\Lambda^{\dagger} \boldsymbol{b}$ by using multinomial expansion theorem (to use multinomial expansion, we must only require $[b_i,b_j]=[b^\dagger_i,b^\dagger_j]=0$ . We first compute $\Lambda$ acting on the input state $\ket{\mathbf l}=\ket{l_1, l_2, \ldots , l_M}$ We have
\begin{align}
      \Lambda\ket{l_1, l_2, . . . , l_M}&=\\
      \prod_{i=1}^{M}\frac{1}{f(l_i)}\sum_{\{ n_{ij}, \sum_{j=1}^{M}n_{ij}=l_i \} } &  \frac{l_i !}{\prod_{i,j=1}^{M}n_{ij}!} (\prod_{j_1=1}^{M} \Lambda_{j_11}b_{j_1}^{\dagger})^{n_{1 j_1}}  (\prod_{j_2=1}^{M} \Lambda_{j_22}b_{j_2}^{\dagger})^{n_{2 j_2}})...(\prod_{j_M=1}^{M}\Lambda_{j_MM}b_{j_M}^{\dagger})^{n_{M j_M}}\ket{0}^M. \\ \nonumber
     & =\sum_{\sum_{j=1}^{M} k_j=N}C_{\mathbf k| \mathbf l}  \ket{\mathbf k}, 
\end{align}
where the coefficient$C_{\mathbf k| \mathbf l}$ is given by  
\begin{equation}
    C_{\mathbf k| \mathbf l}=\prod_{i=1}^{M}\frac{l_i !}{f(l_i)}\prod_{j=1}^{M}f(k_j) \sum_{\{ n_{ij}: \sum_{j=1}^{M}n_{ij}=l_i, \sum_{i=1}^{M}n_{ij}=k_j \} } \frac{\prod_{i,j=1}\Lambda_{ji}^{n_{ij}}}{\prod_{i,j=1}^{M}n_{ij}!}.
\end{equation}
We use the following combinatorial identity~\citep{aaronson2011computational,scheel2004permanents}
\begin{align}
\sum_{\{ n_{ij} \in \mathbb N_0: \sum_{j=1}^{M}n_{ij}=l_i, \sum_{i=1}^{M}n_{ij}=k_j \} } \frac{\prod_{i,j=1}\Lambda_{ji}^{n_{ij}}}{\prod_{i,j=1}^{M}n_{ij}!}=
\left(\prod_{i=1}^{M}\frac{1}{l_i!} \right) \left(\prod_{j=1}^{M}\frac{1}{k_j!}\right)\mathrm{Perm} (\Lambda[\mathbf k| \mathbf l]).
\end{align}
Finally, we get (notice that $f(0)=1$):
\begin{equation}
\label{eq:output boson}
    C_{\mathbf k| \mathbf l}=    \left(\prod_{i=1}^{N} \frac{f(k_i)}{f(l_i) k_i!}\right) \left| \mathrm{Perm} (\Lambda[\mathbf k| \mathbf l]) \right|.
\end{equation}
which finishes the proof as
$\mathrm{Pr}(\mathbf k | \mathbf l)=|C_{\mathbf k| \mathbf l}|^2.$
\end{proof}

In particular, when we choose the standard-bosonic  $f(n)=\sqrt{n!}$, \eqref{eq:output boson} reduces to the output probabilities of  standard FBS
\begin{equation}
\begin{aligned}
   \mathrm{Pr} (\mathbf k| \mathbf l) = 
\frac{
    \left|\mathrm{Perm} (\Lambda[\mathbf k| \mathbf l]) \right|^2 
} {
    \left( \prod_i l_i!\right) \left(\prod_i k_i!\right) 
}   .
\end{aligned}
\end{equation}
which recovers the result of FBS \citep{aaronson2011computational,scheel2004permanents}. 

Since the prefactors of the probabilities will always be a constant in the collision-free subspace, i.e., whenever $k_i, l_i \in \{0, 1\}$, the hardness results of \textcite{aaronson2011computational} directly apply to generalized boson sampling. 



\section{Gaussian Boson Sampling for Generalized Boson}
\label{SM:generalized_GBS}

In this section, we generalize the Gaussian-state formalism to generalized bosons. 
As it turns out, key features of this formalism do not generalize, making it somewhat contrived. 
Nonetheless, it can be used to show that the outcome probability of the analogue of GBS are also given by Hafnians.

To begin with, we introduce the definition of a coherent state, central to the Gaussian formalism, in Sec.~\ref{supp:generalized coherent}. 
We then derive the $P$ and $Q$ function in Sec.~\ref{supp:p q function} and use it to compute the outcome probabilities of generalized GBS in Sec.~\ref{supp:gbs}. 

\subsection{Generalized Coherent States}
\label{supp:generalized coherent}

In this section, we introduce the coherent state of generalized bosons. 
We define $\ket{\alpha}=\frac{1}{\sqrt{N(\alpha)}} e^{\alpha a^{\dagger}}\ket{0}$ for a generalized bosonic operator $a$ with bosonic factor $f$ as 
\begin{equation}
    \ket{\alpha}=\frac{1}{\sqrt{N(\alpha)}}\sum_{n=0}^{\infty} \frac{(\alpha a^{\dagger})^n}{n!}\ket{0}=\frac{1}{\sqrt{N(\alpha)}}\sum_{n=0}^{\infty} \frac{(\alpha)^n}{n!}f(n)\ket{n}, 
\end{equation}
where $N(\alpha)$ is a normalization factor that enforces $\braket{\alpha|\alpha}=1$, as given by 
\begin{equation}
   N(\alpha)=\sum_{n=0}^{\infty}\frac{|\alpha|^{2n}f^2(n)}{n!^2}.
\end{equation}
We can then compute the inner product of two generalized coherent states $\bra{\beta}$ and $\ket{\alpha}$ as 
\begin{equation}\label{eq:generalized_coherent_innprod}
   \braket{\beta|\alpha}=\frac{1}{\sqrt{N(\alpha)N(\beta)}}\sum_{n=0}^{\infty}\frac{(\beta^{*}\alpha)^{2n}f^2(n)}{n!^2}.
\end{equation}
Indeed, if we choose the bosonic factor to be $f(n)=\sqrt{n!}$, then we obtain $N(\alpha)=e^{|\alpha|^2} $, which reproduces the standard coherent state
\begin{equation}
    \ket{\alpha}=e^{-\frac{|\alpha|^2}{2}}\sum_{n=0}^\infty \frac{\alpha^n}{\sqrt{n!}}\ket{n}=e^{-\frac{|\alpha|^2}{2}}e^{\alpha \hat{a}^\dagger} e^{-\alpha^*\hat{a}}\ket{0}.
\end{equation}

\subsection{\texorpdfstring{$P$}{} and \texorpdfstring{$Q$}{} Functions for Generalized Boson}
\label{supp:p q function}

Phase space methods are widely used in quantum optics. 
The advantage of phase space methods is that in order to compute an expectation value of an observable, one only needs to compute an integral over phase space instead of calculating a trace. 
Specifically, the $P$ and $Q$ functions are the most widely used representatios of observables and density matrices in phase space \cite{Garrison_Quantum_Optics,zoller2004quantum}, respectively. 

In this section, we define the $P$ and $Q$ functions for generalized bosons.
Before we start studying the $P$ and $Q$ functions, we need to first figure out the normalization constant of the generalized coherent basis.
In the single mode case, we have the integral
\begin{equation}\label{overcomplete}
    \int \ket{\alpha}\bra{\alpha} d^2 \alpha  = \kappa I.
\end{equation}
where $\kappa$ is a positive constant.
For example, for the $q$-boson coherent state, we have $\kappa=\pi$, and for the spin coherent state, we have $\kappa=\frac{4 \pi}{2S+1}$.
The expectation of an operator $O$ can then be written as an integral over a two-dimensional phase space with coordinates $\alpha, \alpha^*$: 
\begin{equation}
    \tr[\rho O ] = \int d^2 \alpha \frac{1}{\kappa}\bra{\alpha}\rho O\ket{\alpha}.
\end{equation}

We use this expression in order to define the phase space representation of a density matrix $\rho$, named $Q$-function, as $Q_\rho(\alpha)=\frac{1}{\kappa}\bra{\alpha}\rho\ket{\alpha}$ where the $\kappa^{-1}$ factor serves to ensure that the $Q$-function is normalized in the sense of 
\begin{equation}
    \int Q_\rho(\alpha) d^2 \alpha=1\quad \& \quad 
\end{equation} 
Intuitively, we can think of the $Q$-function as the projection of the density matrix $\rho$ onto the coherent-state basis. 
In the $M$-mode case, the $Q$-function is then given by $Q_\rho(\boldsymbol{\alpha})=\kappa^{-M} \bra{\boldsymbol{\alpha}} \rho \ket{\boldsymbol{\alpha}}$, where $\ket{\boldsymbol{\alpha}}=\ket{\alpha_1}\otimes \cdots \otimes \ket{\alpha_M}$.

Since the $Q$-function is the phase space representation of a density matrix $\rho$, similarly, we also map the observable to its corresponding phase space representation namely the $P$-function. 
For the purpose of this work, we only need  observables which are projective measurements in the Fock basis. 
The $P$-function of a single-mode projective Fock measurement $\ket{n} \bra{n}$ is defined as the function $P_n:\mathbb C \rightarrow \mathbb R$ which satisfies
\begin{equation}
    \ket{n}\bra{n}=\int d^2\alpha \ket{\alpha}\bra{\alpha} P_n(\alpha). 
    \label{sec 1:P-representation}
\end{equation}
For the multi-mode case, the $P$-function $P_{\boldsymbol{n}}$ is defined by the analogous relation $\ket{\boldsymbol{n}}\bra{\boldsymbol{n}}=\int d^{2M}\alpha \ket{\boldsymbol{\alpha}}\bra{\boldsymbol{\alpha}} P_{\boldsymbol{n}}(\boldsymbol{\alpha})$.

Given those definitions, we can write the probability of measuring a multi-mode Fock state $\ket{\boldsymbol{n}}=\ket{n_1,n_2,...,n_N}$ on a density matrix $\rho$ as
\begin{equation}\label{particle_number_stat}
\text{Pr}(\boldsymbol{n})=\int d^{2M}\alpha Q_\rho(\boldsymbol{\alpha}) P_{\boldsymbol{n}}(\boldsymbol{\alpha}).
\end{equation}

For generalized bosons, the $Q$-function can be obtained straightforwardly while the $P$-function is nontrivial.
Since $\ket{\boldsymbol{n}}=\ket{n_1}\otimes \cdots \otimes \ket{n_N}$ is a product state, its $P$-function is a product of single-mode $P$-functions acting on each mode individually. 
This can be seen by generalizing  Eq.~\eqref{eq:check p function} to multi-mode Fock states.
It turns out that the single-mode $P$-function for $\ket{n}\bra{n}$ can be written as 
\begin{equation}
    P_{n}(\alpha)=N(\alpha)\frac{1}{f(n)^2}\left(\frac{\partial^2}{\partial \alpha \partial \alpha^*}\right)^{n}\delta(\alpha)\delta(\alpha^*).
    \label{P-c}
\end{equation}
To see this, let us verify it by the definition of $P$-function.  by integrating $P_{n}(\alpha)$ over phase space

\begin{equation}
\label{eq:check p function}
\begin{aligned}
    \int d^2\alpha \ket{\alpha}\bra{\alpha}P_{n}(\alpha)= 
    &\int d^2\alpha \ket{\alpha}\bra{\alpha}N(\alpha)\frac{1}{f(n)^2}\left(\frac{\partial^2}{\partial \alpha \partial \alpha^*}\right)^{n}\delta(\alpha)\delta(\alpha^*)\\
    =&\frac{1}{f(n)^2}\int d^2\alpha\sum_{k,l=0}^\infty \frac{\alpha^k \alpha^{*l}}{k!l!}f(k)f(l)\ket{k}\bra{l}\left(\frac{\partial^2}{\partial \alpha \partial \alpha^*}\right)^{n}\delta(\alpha)\delta(\alpha^*)\\
    =&\frac{1}{f(n)^2}\int d^2\alpha \delta(\alpha)\delta(\alpha^*)\left(\frac{\partial^2}{\partial \alpha \partial \alpha^*}\right)^{n} \sum_{k,l=0}^\infty \frac{\alpha^k \alpha^{*l}}{k!l!}f(k)f(l)\ket{k}\bra{l}\\
    =&\ket{n}\bra{n}.
\end{aligned}
\end{equation}
Here, we use the property of the delta function, since once we take the partial derivative $\left(\frac{\partial^2}{\partial \alpha \partial \alpha^*}\right)^{n}$ inside integral, only the $\ket{n}$ component remains after integration. 
Because the delta function eliminates all the terms depending on $\alpha$ after partial derivatives, $\ket{n}$ is the only component  that does not depend on $\alpha$ after we take the partial derivative. 

The $P$-function of a multi-mode Fock state $\ket{\boldsymbol{n}}$ is then given by 
\begin{equation}\label{generalized_P}
P_{\boldsymbol{n}}(\boldsymbol{\alpha})=\frac{1}{\prod_{i=1}^M f(n_i)^2}e^{\sum_{i=1}^M\ln(N(\alpha_i))} \prod_{j=1}^M \Bigg(\frac{\partial^2}{\partial \alpha_j \partial \alpha_j^* }\Bigg)^{n_j} \delta(\alpha_j)\delta(\alpha_j^*).
\end{equation}

\subsection{Generalized Gaussian states }
\label{supp:gen gauss}

Given the phase-space representation in terms of the $P$ and $Q$ funcction, we define a Gaussian state for generalized bosons analogously to the standard case, if it has a Gaussian $Q$-function in the sense that 
\begin{equation}\label{generalized_Q}
Q_{\rho}(\boldsymbol{\alpha})=\frac{1}{g\sqrt{ |\sigma_{Q}|}}e^{-\frac{1}{2}\boldsymbol \alpha^{\dagger}\sigma_{Q}^{-1}\boldsymbol \alpha} 
\end{equation}
for some positive semidefinite matrix $\sigma_Q$, called the covariance matrix. 
Here, $\boldsymbol \alpha=(\alpha_{1}, \alpha_{2}, ...\alpha_{M})^T, \boldsymbol \alpha ^{\dagger}=(\alpha_{1}^{*}, \alpha_{2}^{*}, ...\alpha_{M}^{*})$, $|\sigma_{Q}|$ is the determinant of $\sigma_{Q}$, and $g$ is a normalization constant which makes $\int Q(\boldsymbol{\alpha}) d^{2M} \boldsymbol \alpha=1$ with $d^{2M} \boldsymbol \alpha=\prod_{i=1}^{M}d^2\alpha_{i}.$

Let us point out, however, that Gaussian states are highly unnatural for generalized bosons. Let us elaborate this point: 
Generally speaking, given a basis, one can define a Gaussian state in the $Q$ representation for generalized bosons. 
Here, we did so by introducing a generalized coherent state, and defining the $Q$-function in the generalized coherent state basis.
However, in contrast to the standard boson, the displacement operator will not prepare a Gaussian state, and neither will the  squeezing operator prepare a Gaussian state.
Hence, it is unclear how to prepare a generalized Gaussian state. 

Put differently, coherent and squeezed states of standard bosons always have a Gaussian $Q$ function, and moreover zero stellar rank within the stellar-rank formalism introduced by \textcite{chabaud2020stellar, chabaud2021classical}. 
Conversely, a coherent or squeezed state of generalized bosons may not to have a Gaussian $Q$ function, the clear classification of states in terms of their stellar rank will not be applicable in general case.   
In particular, the inner product between coherent states is not Gaussian see Eq.~\eqref{eq:generalized_coherent_innprod}.

\subsection{Gaussian Boson Sampling for Generalized Boson}
\label{supp:gbs}

Let us now derive the outcome probability of GBS for generalized boson and show that we still get a Hafnian by assuming our state has a Gaussian $Q$-function.
To do so, we write $P_{\boldsymbol{n}}$ as
\begin{equation}\label{generalized_P}
P_{\boldsymbol{n}}(\boldsymbol{\alpha})=\frac{1}{\prod_{i=1}^M f(n_i)^2}e^{\sum_{i=1}^M\ln(N(\alpha_i))}\prod_{j=1}^M \Bigg(\frac{\partial^2}{\partial \alpha_j \partial \alpha_j^* } \Bigg)^{n_j} \delta(\alpha_j)\delta(\alpha_j^*)
\end{equation}
We then evaluate the phase space integral \eqref{particle_number_stat} with the expressions for the $Q$ function \eqref{generalized_Q}, and the $P$ function~\eqref{generalized_P}, obtaining
\begin{equation}
\text{Pr}(\boldsymbol{n})=\frac{1}{\prod_{i=1}^M f(n_i)^2}\frac{1}{g\sqrt{|\sigma_{Q}|}}\prod_{j=1}^{M} \Bigg(\frac{\partial^2}{\partial \alpha_j \partial \alpha_j^{*} } \Bigg)^{n_j}e^{\ln(N(\alpha)-\frac{1}{2}\boldsymbol \alpha^{\dagger}\sigma_{Q}^{-1}\boldsymbol \alpha}|_{\alpha_j \to 0}.
\end{equation}
From this, we recover standard GBS \cite{hamilton2017gaussian, kruse2019detailed} by choosing $N(\alpha)=e^{|\alpha|^2}$ and $f(n)=\sqrt{n!}$ because the expression in the argument of the exponential can be written as a quadratic form in the matrix $I_{2M} - \sigma_Q^{-1}$, where $I_{2M}$ is the $2M$-dimensional identity matrix.

Indeed, from the perspective of generalized bosons, it looks like for the standard boson case, the resulting expression  is a miracle since we get the exponential quadratic term $e^{|\alpha|^2}$ for the normalization coefficient $N(\alpha)$. 
However, observe that only the second derivative on each mode contributes to the outcome probability if we only measure outcomes $n_j \in \{ 0,1\}$ in every mode. 
This is because when letting $\alpha_j \rightarrow 0 $ the exponential term will approach unity since $N(0) = 1$ for arbitrary generalized bosons. 
This corresponds to the second-order term in the series expansion of $\ln (N(\alpha))$ around $\alpha = 0 $. 
Let us therefore compute the series expansion of $N(\alpha)$ in $|\alpha| = \sqrt{\alpha^*\alpha}$ around $|\alpha|=0$ as 
\begin{equation}
\ln(N(|\alpha|))=c_0+c_1|\alpha|^2+c_2 |\alpha|^4+c_3 |\alpha|^6+ \ldots . 
\end{equation}

As argued above, when restricting to $n_j \in \{ 0,1 \}$ for $1 \leq j \leq M$, only the term linear in $|\alpha|^2 = \alpha^* \alpha$ will give a nontrivial contribution to the outcome probability.
Hence, only $c_1$ is effective in this case, because all the higher order terms in $|\alpha|$ will be eliminated by $\boldsymbol{\alpha}\rightarrow 0$.
We obtain
\begin{equation}\label{eq:output_gGBS}
\text{Pr}(\boldsymbol{n})=\frac{1}{\prod_{i=1}^M f(n_i)^2}\frac{e^{c_0 M}}{g\sqrt{|\sigma_{Q}|}}\prod_{j=1}^{M} \Bigg(\frac{\partial^2}{\partial \alpha_j \partial \alpha_j^{*} } \Bigg)^{n_j}e^{c_1 |\alpha|^2-\frac{1}{2}\boldsymbol \alpha^{\dagger}\sigma_{Q}^{-1}\boldsymbol \alpha}|_{\alpha_j \to 0}.
\end{equation}

Then we do the similar calculations as shown in Refs.~\citep{hamilton2017gaussian, kruse2019detailed} which is based on the derivative expansion formula \cite{hardy2006combinatorics} converting partial derivatives of an exponential quadratic function to a summation over all perfect matching permutations (PMP) of product of matrix elements. Such a summation over PMP of products of matrix elements is exactly the Hafnian function.
We have our final result:
\begin{equation}
\text{Pr}(\boldsymbol{n})= \frac{e^{c_0 M}}{g\prod_{i=1}^M f(n_i)^2} \frac{1}{\sqrt{ |\sigma_{Q}|}} \text{Haf}(A_s)
\end{equation}
Here 
\begin{equation}
A_s=\begin{pmatrix}
  0 & \bigoplus_{i=1}^Mc_1^{(i)} I_M\\ 
  \bigoplus_{i=1}^Mc_1^{(i)} I_m & 0
\end{pmatrix}(I_{2M}-\sigma_{Q}^{-1}).
\end{equation}
We finished the proof.

We also find that parallel results hold for the case of nonzero displacement. 
In this case, the $Q$ function looks like:
\begin{equation}\label{generalized_Q1}
Q_{\rho}(\alpha)=\frac{1}{g\sqrt{ |\sigma_{Q}|}}e^{-\frac{1}{2}(\boldsymbol \alpha-d_{\nu})^{\dagger}\sigma_{Q}^{-1}(\boldsymbol \alpha-d_{\nu})}.
\end{equation}
We calculate the probability to observe a Fock state $\ket{n_1,...,n_M} $ where $n_i\in \{0,1\}$ is \citep{hardy2006combinatorics}.
We arrive at
\begin{equation}
\begin{aligned}
\text{Pr}(\boldsymbol{n})&= 
\frac{ e^{c_0 M}\exp\left[-\frac{1}{2}d_{\nu}^\dag \sigma_Q^{-1} d_{\nu}\right]}{g\prod_{i=1}^M f(n_i)^2\sqrt{|\sigma_Q|}} \left. \prod^{M}_{j=1} \left(\frac{\partial^2}{\partial \alpha_j \partial \alpha_j^*}\right)^{n_j}\hspace{-3mm}\exp\left[\frac{1}{2}\boldsymbol \alpha^t A \boldsymbol \alpha+ F \boldsymbol \alpha\right]\right|_{\mathbf\alpha=0} \, .
\end{aligned}
\end{equation}
We get:
\begin{equation}
\begin{aligned}
\text{Pr}(\boldsymbol{n}) &=\frac{e^{c_0 M} e^{-\frac{1}{2}d_{\nu}^\dag \sigma_Q^{-1} d_{\nu}}}{g\prod_{i=1}^M f(n_i)^2\sqrt{|\sigma_Q|}} \sum^{|\pi |}_{\substack{j=1 \\ \pi_j \in \{2M\} }} \left[  \left ( \prod^{|B^1_j|}_{\substack{k =1 \\ B_j^1 \in  \pi_j}} F_k \right ) \mathrm{Haf}(A_{B_j^2}) \right]
\\
 &=\frac{e^{c_0 M} e^{-\frac{1}{2}d_{\nu}^\dag \sigma_Q^{-1} d_{\nu}}}{g\prod_{i=1}^M f(n_i)^2\sqrt{|\sigma_Q|}} \Bigg [ \mathrm{Haf}(A_S)  + \sum_{j_1,j_2 ,j_1\ne j_2} F_{j_1} F_{j_2} \mathrm{Haf}(A_{S - \{j_1,j_2\}}) + ... + \prod^{2M}_j F_j \Bigg ]  \\
\end{aligned}
\label{eq:squ_coh_light}
\end{equation}
where the first sum is over all partitions of the set  of $2M$ indices , the product is over all indices in the blocks $B^1_j$ and the remaining indices in blocks $B^2_j$ form $A_{B^2_j}$, a submatrix of $A$, which we then take the hafnian of. 

\section{Trapped-ion Implementation
\label{SM:trapped_ion}}

In this section we discuss the details of physical implementation
of the boson sampling protocol in trapped-ion setups.

\subsection{Setup\label{subsec:Setup}}

The setup we have in mind is shown in Fig.~1 of the main text. More
precisely we consider a chain of ions in a Paul trap. Each ion is
considered to be a two-level system $\left\{ \left|g\right\rangle _{i},\left|e\right\rangle _{i}\right\} $
with the corresponding transition frequency $\omega_{eg}$. We encode
superspin $S^{(\alpha)}$ as collective excitation of a subset of
ions $\left\{ i\right\} _{\alpha}$ such that the lowering operator
can be defined as:

\begin{align}
S_{\pm}^{\left(\alpha\right)} & \equiv\frac{1}{\sqrt{N}}\sum_{\{i\}_{\alpha}}\sigma_{\pm}^{(i)},\label{eq:S_p}
\end{align}
where $N$ is the number of ions encoding the superspin and $\sigma_{-}^{(i)}\equiv\left|g\right\rangle _{i}\left\langle e\right|$,
$\sigma_{+}^{(i)}\equiv\left|e\right\rangle _{i}\left\langle g\right|$.

\subsubsection*{M{ø}lmer-Sorensen interaction}

As extensively discussed in the Ref.~\cite{brydges2019probing},
the Ising-type interaction between of ions can be generated using
the Moelmer-Sorensen \cite{molmer1999multiparticle} laser configuration.
More precisely we consider a pair of bichromatic laser beams driving
the transition $e\leftrightarrow g$ with the laser frequencies respectively
tuned to $\omega_{eg}+\Delta$ and $\omega_{eg}-\Delta$ where $\Delta$
is some detuning having two different Rabi frequencies $\Omega_{\pm}$.
Here we assume the possibility of selective driving of a pair of ions
as shown in Fig.~1. This generates the ion-ion interaction
characterized by the Ising Hamiltonian:

\begin{equation}
H_{\text{I}}=J_{i,j}\sigma_{x}^{(i)}\sigma_{x}^{(j)}+h\left(\sigma_{z}^{(i)}+\sigma_{z}^{(j)}\right),\label{eq:H_I}
\end{equation}
where the interaction coefficient scales as $J_{ij}=J_{0}/\left|i-j\right|^{\zeta}$
with $0<\zeta<3$ and $h$ denotes the transverse field.

\subsection{Mode-mixing operation\label{subsec:Mode-mixing-operation}}

In this section, we discuss the implementation of the two-mode mixing
operation on a pair of superspins. More precisely we show how an interaction
with the Hamiltonian $H_{\text{int}}=J^{(\alpha,\beta)}S_{+}^{(\alpha)}S_{-}^{(\beta)}+\text{H.c. }$
can be achieved in the setup described above. As discussed and experimentally
demonstrated in \cite{brydges2019probing} the basic XY Hamiltonian
between any two spins can be achieved from the Ising Hamiltonian Eq.~\eqref{eq:H_I}
in the limit of a large transverse field. In this case only the excitation-number
preserving terms remain relevant and we find:

\begin{align}
H_{i,j}^{(XY)} & =J_{i,j}\left\{ \sigma_{+}^{(i)}\sigma_{-}^{(j)}+\text{H.c.}\right\} +h\left(\sigma_{z}^{(i)}+\sigma_{z}^{(j)}\right).\label{eq:HXY}
\end{align}
Since the interaction preserves the number of excitations we can absorb
the transverse-field term by transforming into the interaction picture
with respect to it. The desired Hamiltonian $H_{\text{int}}$ can
be obtained in a Trotterized way by running the interaction Eq.~\eqref{eq:HXY}
for each pair of ions for a time $\delta t_{i,j}$ inverse proportional
to the interaction strength $J_{i,j}$ to preserve the spin symmetry
Eq.~\eqref{eq:S_p} $\delta t_{i,j}^{\alpha,\beta}\approx\delta t_{\alpha\beta}\times\tilde{J}/J_{i,j}$,
where $\tilde{J}=\text{min}_{\{i\}_{\alpha}\{j\}_{\beta}}\left[J_{i,j}\right]$.
To the lowest-order in Floquet-Magnus expansion we find:

\begin{align*}
H_{\text{eff}} & \approx\frac{\sum_{i,j}\delta t_{i,j}J_{i,j}\left\{ \sigma_{+}^{(i)}\sigma_{-}^{(j)}+\text{H.c.}\right\} }{\sum_{i,j}\delta t_{i,j}}\\
 & \approx J_{0}N\frac{\sum_{\alpha,\beta}\delta t_{\alpha\beta}\times\left\{ S_{+}^{(\alpha)}S_{-}^{(\beta)}+\text{H.c.}\right\} }{\sum_{\alpha,\beta}\sum_{i,j}\delta t_{\alpha\beta}\times\left|i-j\right|^{\zeta}}
\end{align*}
This Hamiltonian is equivalent to the Hamiltonian of interest $H_{\text{int}}.$
We can now estimate the order of magnitude of the interaction strength
for the nearest-neighbor superspins. Using \cite{Jurcevic:2014te,brydges2019probing}
$\zeta\approx1$ and $J_{0}=1\text{kHz}$ we find the overall interaction
constant for the nearest and next-nearest neighbor superspins being
respectively $J^{(\alpha,\alpha+1)}\approx44\text{Hz}$ and $J^{(\alpha,\alpha+2)}\approx21\text{Hz}.$
For comparison the typical decoherence rates can be estimated to be
of the order of 1Hz \cite{brydges2019probing}. 

\section{Circuit-QED implementation}
\label{SM:circuit_qed}

In this section we provide details of the circuit-QED implementation
of the boson-pair generalized boson. The preparation step is discussed
in the main text. Here we only focus on the implementation of the
mode-mixing unitary operation. In this derivation we closely follow
\cite{PhysRevLett.123.063602}.

\subsection{Mode-mixing Hamiltonian}

We now consider the implementation of the mode-mixing unitary operation.
Here for simplicity we only consider two cavities ($\omega_{i=1,2}$
denote frequency and $a_{i}$ is the mode annihilation operator) which
interact with four Josephson junctions ($\omega_{J}$ denote frequency
and $J$ is the mode annihilation operator) and two two-level systems. The Hamiltonian of the
system reads \cite{PhysRevLett.123.063602} $H=H_{0}+H_{d}$:

\begin{align}
H_{0}= & \omega_{c}\sum_{\lambda}a_{\lambda}^{\dagger}a_{\lambda}+\omega_{J}\sum_{\lambda}J_{\lambda}^{\dagger}J_{\lambda}+\omega_{0}c^{\dagger}c-\frac{E_{J}}{24}\sum_{i}\phi_{i}^{4},\label{eq:H0}\\
H_{d}  = & \sum_{i}\Omega_{i}\left(t\right)J_{i}^{\dagger}+\text{H.c.,}\label{eq:Hd}
\end{align}
where the phases are:

\begin{align*}
\phi_{1} & =\phi_{a}a_{1}+\phi_{b}b+\phi_{J}J_{1}+\text{H.c.,}\\
\phi_{2} & =\phi_{a}a_{2}+\phi_{b}b+\phi_{J}J_{2}+\text{H.c.,}
\end{align*}
and $\phi_{a,b,J}$ denote the corresponding participation ratios.
The Rabi frequencies are assumed to be given by:
\[
\Omega_{i}\left(t\right)=\Omega_{i}e^{-i\omega_{d}t}
\]
We now transform into the interaction frame with respect to the driving
frequency $\omega_{d}$ and discard the rapidly rotating terms. Renormalizing
the frequencies such as to include the Stark shifts due to driving
the effective Hamiltonian can be put under the following form (we
ignore quartic terms for $b$ as we assume it does not contain more
than 1 photon at the time):

\begin{align*}
H_{\text{eff}} & =\sum_{i}\left(g_{i}a_{i}^{\dagger2}c+\text{H.c.}\right)-\Delta c^{\dagger}c-\chi\sum_{i}a_{i}^{\dagger}a_{i}^{\dagger}a_{i}a_{i},
\end{align*}
where $g_{i}=-\frac{1}{2}\phi_{b}\phi_{c}^{2}\phi_{J}\frac{\Omega_{i}}{\Delta_{J}}$
is a complex effective tunneling coefficient, $\Delta_{J}=\omega_{d}-\omega_{J}$, 
and the induced quartic non-linearity is $\chi=\frac{\phi_{c}^{4}}{4}$.
The detuning is denoted as $\Delta=2\omega_{a}-\omega_{0}-\omega_{d}$.
In the following we will drop the on-site nonlinear term proportional
to $\chi$. As discussed in \cite{PhysRevLett.123.063602} it can
be dynamically compensated by coupling to an additional Josephson
qubit. 

\subsubsection{Adiabatic elimination of cavity bus}

We now perform the adiabatic elimination of the cavity bus degree
of freedom $c$ assuming the driving is weak enough such that $g_{i}\ll\Delta$.
The resulting Hamiltonian is 

\begin{align*}
H_{\text{eff}} & =\sum_{i,j}\frac{g_{i}g_{j}^{*}}{\Delta}a_{i}^{\dagger2}a_{j}^{2}.
\end{align*}
Now assuming the coupling coefficients are time-dependent $g_{i}=g_{i}\left(t\right)$
we can implement any coupling between any pair of sites in a Trotterized
fashion in complete analogy to Sec.~\ref{subsec:Mode-mixing-operation}. 

\subsection{Fock-state preparation}

We now consider the preparation of Fock state $\frac{1}{\sqrt{2}}a_{i}^{2}\left|0\right\rangle $
in the system described by the Hamiltonian (\ref{eq:H0}, \ref{eq:Hd})
but with the driving Hamiltonian to be of the form 
\[
\ensuremath{H_{\text{d}}=\Omega_{\text{dr}}\sum_{i}\cos(\omega_{\text{dr}}t)(a_{i}+a_{i}^{\dagger})}.
\]
In rotating-wave approximation we get for each site:

\[
H_{\text{eff}}=\frac{\Omega_{\text{dr}}}{2}(a_{i}+a_{i}^{\dagger})-\Delta_{a}a_{i}^{\dagger}a_{i}-\chi a_{i}^{\dagger}a_{i}^{\dagger}a_{i}a_{i}
\]
with $\Delta_{a}=\omega_{\text{dr}}-\omega_{a}$. We now assume the
system is initially prepares in the Fock $\left|0\right\rangle $
state and the frequency of the drive is tuned into resonance with
the two-photon resonance such that $\Delta_{a}=-\chi$. The system
undergoes Rabi oscillations $\left|0\right\rangle \rightarrow\left|2\right\rangle $
with the Rabi frequency given by $\ensuremath{\Omega_{2}=\sqrt{2}\Omega_{\text{dr}}^{2}/(4\chi)}$.
The estimate of fidelity can be obtained by comparing the Rabi frequency and the corresponding detunings as discussed in the main text.


\bibliography{references}